\documentclass[preprint2]{aastex}
\usepackage{natbib}
\usepackage{graphicx}

\shorttitle{Disc Masses of SINGS Galaxies}
\shortauthors{de~Denus-Baillargeon et al.}

\begin{document}

\title{A galactic weigh-in:  mass models of SINGS galaxies using chemospectrophotometric galactic evolution models}
\author{M.-M.~de~Denus-Baillargeon and O. Hernandez}
\affil{Laboratoire d'astrophysique exp\'erimentale, D\'epartement de physique, Universit\'e de Montr\'eal, C.P. 6128,  succursale Centre-ville, Montréal (Qc), H3C 3J7, Canada}
\author{S. Boissier and P. Amram}

\affil{Universit\'e Aix-Marseille, CNRS, LAM (Laboratoire d'Astrophysique de Marseille), UMR 7326, 13388 Marseille, France}
\and 
\author{C. Carignan}
\affil{Dept. of Astronomy, University of Cape Town, Rondebosch 7701, South Africa\\Laboratoire d'astrophysique exp\'erimentale, D\'epartement de physique, Universit\'e de Montr\'eal, C.P. 6128,  succursale Centre-ville, Montréal (Qc), H3C 3J7, Canada\\Observatoire d'Astrophysique de l'Universit\'e de Ouagadougou, BP7021, Ouagadougou 03, Burkina Faso}

\begin{abstract}The baryonic mass-to-light ratio ($\Upsilon_{\star}$) used to perform the photometry-to-mass conversion has a tremendous influence on the measurement of the baryonic content and distribution, as well as on the determination of the dark halo parameters.  Since numerous clues hint at an inside-out formation process for galaxies, a radius-dependant $\Upsilon_{\star}$ is needed to physically represent the radially varying stellar population. 
In this article, we use chemo-spectrophotometric galactic evolution (CSPE) models to determine $\Upsilon_{\star}$ for a wide range of masses and sizes in the scenario of an inside-out formation process by gas accretion. We apply our method on a SINGS subsample of ten spiral and dwarf galaxies for stellar bands covering from the UV to the MIR. The CSPE models prove to be a good tool to weight the different photometric bands in order to obtain consistent stellar discs' masses regardless of the spectral band used. On the other hand, we show that colour index vs. $\Upsilon_{\star}$ relation is an imperfect tool to assign masses to young stellar populations because of the degeneracy affecting $\Upsilon_{\star}$ in all bands at low colour index. Resulting discs from our analysis are compatible with the maximum disc hypothesis provided that adequate bulge/disc decomposition is performed and correction for the presence of a bar is not neglected since it disturbs the internal disc kinematics. Disc-mass models including $\Upsilon_{\star}$-as a free parameter as well as models using our physically motivated radial variation of $\Upsilon_{\star}$ are presented and discussed for each galaxy.
\end{abstract}

\keywords{galaxies: individual (SINGS), galaxies: kinematics and dynamics, galaxies: stellar contents}

\section{Introduction}

The question of the exact contribution of the stellar disc to the overall galaxy kinematics is a long-acknowledged problem that has received a lot of attention for as long as the mass models themselves, starting with the first determination of the mass of the Andromeda galaxy. \citet{Oepik1922} then used its rotational velocity and assumed that its stellar mass is proportional to its total luminous emission. The introduction of inhomogeneous ellipsoids to account for the different galactic populations by \citet{Perek1948,Kuzmin1952,Schmidt1956} allowed modelling of the radial mass distribution. A step forward was taken when \citet{deVauc1953} established from photometric observations that the stellar surface density decreases exponentially with the radius. Despite the consequently expected keplerian decrease of the circular velocity with radius pioneer papers found a decrease much lower than keplerian \citep{Rogstad1970,Roberts1975,Freeman70,Rubin1978}. 

The extended flat HI rotation curves available only from the very late seventies led to the introduction of an additional component to model the rotation curve: the dark halo. This completely changed our vision of the mass distribution in spiral galaxies.  
The total radial mass distribution of spiral galaxies is broken up into several components: a disc consisting mostly of stars depending on a free parameter, the disc mass-to-light ratio ($\Upsilon_{\star}$), plus eventually a bulge depending on another free parameter (the bulge mass-to-light ratio $\Upsilon_B{\star}$); an HI+He disc which does not contain any free parameter (the helium fraction not being a free parameter) and a halo, generally spheroidal, which contains the dark matter and should be described by (at least) two free parameters.  In order to convert the surface brightness photometry of a galactic disc into a radial density profile, an estimation of $\Upsilon_{\star}$ has to be assumed to account for the disc being composed of billions of stars of masses, ages and metallicities different from our Sun's. This directly raises the question of disc-halo degeneracy that consists in balancing the respective contributions of the disc and halo.  Using \textit{ad hoc} parameters, the disc mass-to-light ratio might range anywhere from $\Upsilon_{\star}\simeq 0$ (the minimum disc hypothesis valuable for low surface brightness spirals or dwarfs) to $\Upsilon_{\star}=\Upsilon_{\star}^{Max}$ \citep[the maximum disc hypothesis applicable to bright early type spirals, e.g.][]{vanAlbada1985}).  The solutions that maximized the disc were favoured by earlier authors \citep{CF85,Bahcall1985,Kent1986}.

In the nineties, the first N-body simulations taking only the dark matter into account became available and they suggested that the dark halo density profiles were peaked in the inner region of the galaxies \citep[cuspy distribution, e.g.][]{NFW96}, while the observations showed the opposite \citep[core distribution, e.g.][]{Blaisous99,deBlok2001}.  More recent simulations reveal that the slope keeps getting shallower towards small radii \citep{Navarro2004,Hayashi2004}. In the mean time, different authors \citep[e.g.][]{Blaisous2004, Dutton2006, Bershady2010a} have shown that the indetermination arising from the photometry-to-mass conversion of the stellar disc can lead to differences of up to a factor of 20 in the mass of the dark halo thus inferred, and hampers the fixing of the exact shape of density profile of the dark matter halo (e.g. cusp vs. core controversy).  

Our poor understanding of the baryonic processes involved in galaxy formation probably leads to this inconsistency between the predictions of the $\Lambda$CDM theory and the observations. Physical processes like adiabatic compression have been invoked to contribute even further to the cuspy dark matter distribution. The dissipation of the disc, via infall of baryons, is thought to compress the dark halo distribution through adiabatic contraction \citep{Blumenthal1986}. Baryonic infall increases rotation velocity in the inner regions, thus the effect of adiabatic contraction of the halo by the disc is to steepen even more the cuspy distribution. 
On the other hand, without major change to the $\Lambda$CDM scenario, numerous authors introduced physical processes that might turn a cusp into a core-like feedback, dynamical friction, merging, spin segregation, halo triaxiality, bar-driven evolution, all effects that could reconcile the simulations with the observations. The effects of bars could be to radially redistribute the baryonic matter \citep{Weinberg2002} .  Merging of cored dark matter haloes might change the dark matter distribution \citep{BKM2004,Dehnen2005}. The dynamical friction of initially very steep cusp heated by subhaloes can convert them in shallower distributions \citep{RD2009}. Feedback might be responsible for baryonic blowouts and baryonic mass redistribution \citep{NFW96,Burkert1995,Gelato1999}. N-body + hydrodynamical simulations assuming the presence of CDM and a cosmological constant are now able to produce less steep dark matter density profile within the central kpc of dwarf galaxies in introducing strong outflows from supernovae which inhibits the bulge formation \citep{Governato2010,Maccio2012,Governato2012}.

If the mass of the disc was to be realistically calculated with the help of models based upon physical motivations, one over the three (or more) free parameters could be fixed and the task at hand reduced to the determination of the shape of the dark halo. The mass-to-light ratio $\Upsilon_{\star}$ may be constrained using arguments based on: dynamics (spiral structure and swing amplification, the flaring of the HI disc, bar formation, gas flow in disc along bars or spirals, velocity dispersion in face-on and edge on galaxies), stellar populations (colour-$\Upsilon_{\star 0}$ relation), deviations from the Tully-Fischer relation and lensing. Unfortunately the problem is far from being unambiguously constrained by these different methods and they lead to different results.

In this paper we will focus on constraints from stellar populations by considering the evolution of the galactic stellar components and determining the collective properties of sets of stars \citep{BdJ00,BdJ01,BC03}.  This approach is very promising since it should be fit to distinguish between all particular cases of stellar populations. Stellar populations differ amongst galaxies but also within them, numerous smoking guns such as radial differences in colours and metallicities pointing to the same direction: an inside-out formation process \citep{Grebel2010}. Using a constant $\Upsilon_{\star}$ would be equivalent as assuming a uniform stellar population throughout the galaxy. Any approach that would want to give a realistic weight to a stellar disc should take into account this radial variation and not only a global value. Some work has already been performed with that intent. Interesting work on this matter was carried out by \citet{Portinari2010} who analyzed the effects of radially varying $\Upsilon_{\star}$ on mass modelling of toy galaxies. \citet{Walter2008}, \citet{Kassin2006} have also achieved interesting results with their use of colour-$\Upsilon_{\star}$ to fix the contribution of their galactic discs to the overall mass distribution. We use in this paper galactic chemo-spectrophotometric evolution (CSPE) models to derive the $\Upsilon_{\star}$ for each galaxy of a SINGS subsample in up to twelve photometric bands.  With the help of this stellar mass-to-light ratio, we then propose a realistic mass model including a disc compounded from photometric observations in wavelengths ranging from the UV to the mid-IR. Details of the CSPE models and methodology of the transformation of disc's photometry into corresponding mass is described in the following section (section \ref{sect:Meth}). \\

The relation between the global characteristics of the CSPE models and the $\Upsilon_{\star}$ they produce are presented in the first part of section \ref{sect:Results} while the second part reports on the detailed mass modelling of NGC 2403 along with the main conclusions gathered from performing the same operation over nine other galaxies of the SINGS sample. We then analyse further our results in section \ref{sect:Discussion} and discuss the limitations of our method. Results of the exact mass modelling of individual galaxies are presented as an appendix.\\

\section{Methods}

\subsection{Stellar disc evolution models}
\label{sect:Meth}

Stellar surface density profiles are computed on the basis of full CSPE models described in detail in \citet[hereafter BP99 and BP00]{BP99,BP00}, with the most recent update and application to the SINGS galaxies in \citet{JCM2011}.  These models will thus not be described in detail here; only a broad outline of them will be given in the following paragraph. 

In the CSPE models, the chemical evolution is computed for each galaxy in concentric rings evolving independently.  An infall of primordial gas is assumed, and a radial as well as temporal normalization is performed to account for different accretion histories of individual galaxies and an inside-out formation scheme. 

The implementation of the star formation rate $\Psi$(SFR) in the models is inspired from \citet{Kennicutt1998} and \citet{Wyse1989}.
It depends on the local gas density (usual Schmidt Law) and angular velocity that may be due to 
the spiral arms frequency or to dynamical aspects \citep[see e.g.][]{Boissier2013}. The angular velocity input in the models is computed from a baryonic disc profile embedded in a pseudo-isothermal sphere. The newly formed stars are distributed along a multi-slope Kroupa-type power-law initial mass function (IMF).  Two variants of this IMF are considered: \citep[equation 2 of ][hereafter K01 and KTG93]{K01,KTG93}.
 It has been verified that the exact shape of the input rotation curve, i.e. using a NFW dark halo profile or even an experimental rotation curve does not affect strongly the overall chemical evolution.  It is the absolute value of $v_c$ that impacts the most the results, the slight radial variations in the input velocities being meaningless compared to the uncertainties related to other ingredients of the models such as star formation efficiency, yields of various chemical elements, etc.  

Once the chemical evolution of the galaxy is solved, the spectrophotometric properties are computed using the Geneva group stellar evolution tracks \citep{Charbonnel1996} and the Lejeune stellar spectra library
\citep{Lejeune1997}, 
all being metallicity-dependent.

Assuming our own galaxy is typical, the model was calibrated using properties of the MW such as the local SFR, stellar and gas surface density, the disc scale-length, the abundance gradient, the stellar and gas profiles and the metallicity distribution of G-dwarfs \citep{BP99}.  

The model was generalized to all disc galaxies in BP00, following
a cosmological framework of galaxy formation \citep{FallEfsthathiou80,MMW98}. 
This context offers scaling relations with respect to those of the MW allowing to relate the 
disc properties to the dark matter halo in which the baryonic disc resides. A grid of models was thus built by varying the two parameters $v_c$ and $\lambda$ of a pseudo-isothermal sphere halo, where $v_c$ is the maximal circular velocity of the disc and $\lambda$ its spin parameter. 

While the description above concerns the construction of a grid of theoretical models
for the evolution of galaxies, the assignation of a given model to an observed galaxy
can be made e.g. on the basis of multi-wavelength profiles (corrected for extinction).
A $\chi^2$ best-fit procedure was performed by \citet[hereafter JCM11]{JCM2011} to find the model best representing the photometry of the SINGS galaxies amongst a grid of simulations with $\lambda = [0.020;0.090]$  and $v_c = [80;360]$.   Using the preliminary results of the $\chi^2$ fitting procedure of JCM11, as well as our own fitting procedure for some of the galaxies of our sample, we linearly interpolated the original grid to calculate all the physical quantities required for the calculus of the stellar mass to light ratio $\Upsilon_{\star}$.  The final haloes grid was refined to $\Delta \lambda = 0.001$, $\Delta v_c = 1$.  We used our own fitting procedure when the adopted distances were different from JCM11 and for galaxies that required breaking up in bulge and disc components.\\ 

Once a best model is chosen, it provides a mass-to-light ratio $\Upsilon_{\star}$ varying smoothly with radius. This $\Upsilon_{\star}$ is then interpolated to the radii of the observed galaxy and is used along with its luminosity to obtain its detailed surface density. The solar magnitudes used in this work to convert surface brightness of the galaxies into solar luminosities were drawn from \citet{Oh2008} for the IRAC bands, http://mips.as.arizona.edu/$\sim$cnaw/sun.html for the FUV and NUV bands and \citet{BlantonRoweis2007} for all the other bands in visible and near-IR. 

The new disc's surface density obtained by this method is converted to effective rotational velocity with the help of the task \textit{rotmod} from the \textsc{Gipsy} package \citep{vanderHulst1992,Vogelaar2001}. Such discs will be referred to as "weighted discs" in the remainder of this paper.  The disc adopted for mass modelling is the median surface density of discs in all bands ranging from the UV to the IR and then converted with \textit{rotmod}. The highest and lowest surface densities at each radius are adopted as the upper and lower boundaries of the error on the disc. 
 
Once the contribution of the stellar disc to the rotation curve has been fixed 
in this manner, it is possible to perform a mass modelling of the rotation 
curve with usual methods, but with a lower degree of freedom,
and a physically motivated stellar disc.

\subsection{The sample}

In order to test the method presented in \ref{sect:Meth}, we applied it
to the galaxies listed in table \ref{tab:Sample}.  
This list is a sub-set of the SINGS sample, a sample that no longer needs long introduction \citep{Kennicutt2003}. 
Its galaxies were chosen to cover the range of properties observed in nearby galaxies.
The high-quality data gathered for the sample has been the object of an abundance of publications and is still the base for numerous projects \citep{Walter2008,JCM2009a,Dale2011}.  We skimmed through the sample by applying three criteria: 1) as late-type as possible to avoid very prominent bulges 2) good quality of fit of the models to the photometry 3) inclination allowing for accurate determination of rotational velocity.

\begin{deluxetable}{lccccccp{4.25cm}}
\tabletypesize{\scriptsize}
\tablecaption{\label{tab:Sample}SINGS Subsample used in this tudy}
\tablewidth{0pt}
\tablehead{
\colhead{Name}	&\colhead{Type\tablenotemark{a}} 	&\colhead{Distance\tablenotemark{b}} &\colhead{P.A} &\colhead{Inclination}	&\multicolumn{2}{c}{Model parameters} 	&\colhead{Distinctive feature}\\
	&	&\colhead{Mpc}	&\colhead{($^{\textsf{o}}$)}	&\colhead{($^{\textsf{o}}$)}	&\colhead{$\lambda~(10^{-2})$} 	&\colhead{$v_c~(kms^{-1}$)}	&
}
\startdata
		NGC 0925				&SAB(s)d	   	&9.3		&282 	&58	 	&9.5		&152	&Presence of a bar\\
		NGC 2403 				&SAB(s)cd	&3.2		&307 	&58	 	&5.1		&116	&Presence of a very small bar\\
		NGC 3198				&SB(rs)c	    	&13.8	&35 		&70	 	&4.7		&172	&Presence of a bar\\
		NGC 3621				&SA(s)d	     	&6.6		&339 	&57	 	&2.5		&146	&\\
		NGC 4254				&SA(s)c		&17		&35 		&30	 	&2.8		&240	&\tablenotemark{c,e}\\
		NGC 4321				&SAB(s)bc	&18		&30 		&32	 	&4.0		&293	&Prominent bulge and presence of a bar\tablenotemark{d,e}\\
		NGC 4569				&SAB(rs)ab	&17		&23 		&65		&4.0		&270&Prominent bulge and presence of a bar; gas depleted Virgo cluster galaxy; starburst in the central kpc, LINER\tablenotemark{c,e,f}\\
		NGC 5055				&SA(rs)bc		&10.1	&285 	&57	 	&3.0		&250	&\\
		NGC 7793				&Sa(s)d		&3.9		&278 	&49	 	&3.9		&101	&\\
		DDO 154					&IB(s)m		&4.3		&35 		&44 		&7.7		&32		&Dwarf galaxy\\
	\enddata

\tablenotetext{a} {Classification are from the NED database  \citep[RC3 catalog,][]{deVauc63}}
\tablenotetext{b} {Distances are those adopted for the SINGS sample as in the works by \citet{Walter2008,GildePaz2007,Kennicutt2003}}
\tablenotetext{c} {HI velocities from \citet{Guhathakurta1988}}
\tablenotetext{d} {HI velocities from \citet{Knapen1993}}
\tablenotetext{e} {HI surface densities from \citet{Chung2009}}
\tablenotetext{f} {Special model for the evolution history of this particular galaxy}

\end{deluxetable}

The photometric data for those galaxies come from archives of the GALEX, 2MASS and SDSS surveys as reduced and corrected for extinction and published in \citet{JCM2009a} and \citet{JCM2009b}, hereafter JCM09a,b.  
 The kinematics data come from the SINGS H$\alpha$ and THINGS \textsc{Hi} studies respectively, except otherwise mentioned in table \ref{tab:Sample}. Reduction processes are presented in \citet{Walter2008}, \citet{Daigle2006} and  \citet{Dicaire2008}. \\

\section{Results}\label{sect:Results}

\subsection{$\Upsilon_{\star}$ of model galaxies}
	\label{sect:UPS}

\begin{figure}
\begin{center}
\epsscale{1}
\plotone{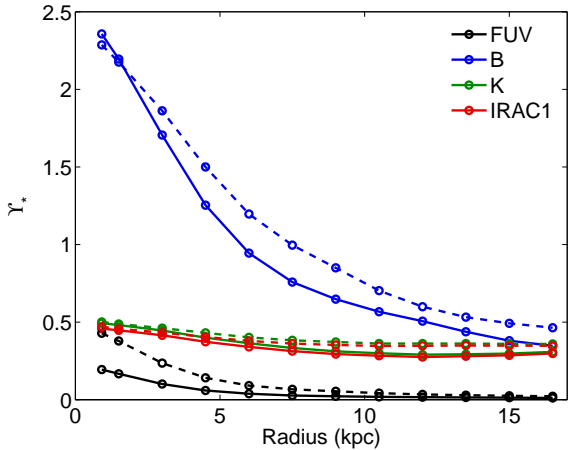}
\caption{\label{fig:CompIMF}Comparison of the $\Upsilon_{\star}$ profile in NUV, B, K and IRAC1 bands for a model matching the parameters of the Milky Way:  $\lambda = 0.03$ and $v_c = 220 km s^{-1}$  The model generated with a KTG93 IMF is represented in a dashed curve and in solid line is the K01 model.} 
\end{center}
\end{figure}

Figure \ref{fig:CompIMF} compares the $\Upsilon_{\star}$ of models including either KTG93 or K01 IMFs for fixed scaling parameters ($\lambda = 0.03$ and $v_c = 220$) and shows some slight differences in the radial behaviour and absolute value of $\Upsilon_{\star}$ in one IMF compared to the other, especially in the UV bands.

This effect is due to the shallower top-heavy end in the K01 IMF by which a larger number of massive stars are produced.  Despite this difference, the results of the two IMFs are very close to one another.  This is due to the K01 models having a smaller fraction of mass trapped in the low mass remnants. That increases the amount of gas available for star formation, which in turn leads to a higher SFR through the evolution of the galaxy, compensating partially for the lower amount of stellar mass locked by generation. As has already been pointed out by \citet{JCM2009a}, the K01 IMF tends to slightly overestimate the UV fluxes of early-types spirals.  Nevertheless, we will show in the following section that K01 models give better fits to the photometric data and more consistent results for weighting the stellar disc in all bands than KTG93 models.  

\begin{figure}
\begin{center}
\epsscale{1}
\plotone{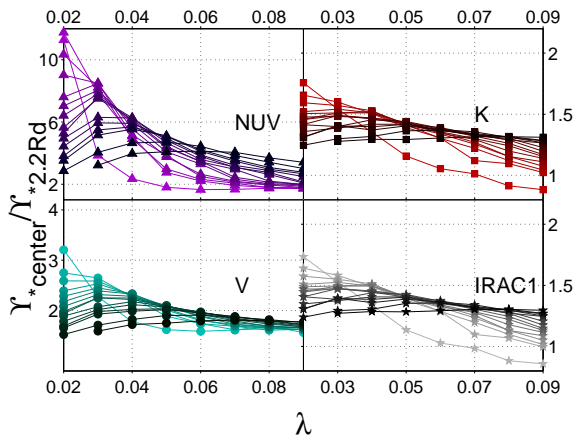}
\caption{\label{fig:DeltaUps}   Relation between the variation of $\Upsilon_{\star}$ from centre to $2.2R_d$ and parameters $\lambda$ and $v_c$. Series of markers joined by a line have a common $v_c$. Higher $v_c$ ($v_c$ = 360) are represented with the darkest shades.}
\end{center}
\end{figure}

The appearance of the $\Upsilon_{\star}$ profiles (fig. \ref{fig:CompIMF}) is the direct consequence of the inside-out galaxy formation process:  the $\Upsilon_{\star}$ is at its maximum at the centre of the galaxy and then decreases to the exterior due to the progressively younger stellar population and lower metallicities, which was shown for late-type spirals even when the colour gradients are relatively small \citep{Carignan85}.   This tendency of increasing $\Upsilon_{\star}$ with age suffers one exception, though, from the infrared bands for a few parameters $\lambda$ and $v_c$ of the scaling haloes because evolved stellar stages, though transient, contribute enormously to the overall luminosity of a stellar population \citep{CB91}.  Larger variations are found at shorter wavelengths since their light is increasingly dominated by short-lived stars \citep{BC03,BP99} that are found in larger proportion in galaxies outskirts. This is remarkable in the UV bands where a more than tenfold variation can be seen on figure \ref{fig:DeltaUps} from centre to the $R_{2.2d}$ radius.\\

The use of $R_{2.2d}$ as determined in the IRAC1 bands links the variation of the $\Upsilon_{\star}$ to the underlying mass distribution rather than only the luminosity.   Let us stress however that the models assume a continuous star formation. It thus does not represent an accurate picture of a star formation history that can be somewhat more eventful in particular galaxies and since it ignores the contribution of bulges and bars it passes over mechanisms that can influence greatly the exact composition of stellar populations \citep{KK04}. \\

The scaling of the properties by different sizes and concentration factors of haloes influences the star forming histories of the resulting model galaxies by changing the accretion rate and velocities of the model galaxy, the SFR at a given radius depending directly on those two factors.\\

In JCM11, the authors point out in their figure 7 the relation between morphological type and the parameters of the scaling halo. As could be expected, the circular velocity differs with type, peaking for Sc galaxies.   The $\lambda$ parameter, on the other hand, shows no tendency whatsoever in the distribution except for maybe a larger dispersion for extremely late type galaxy. They stress that this tends to confirm predictions of the $\Lambda$CDM paradigm that the angular momentum per unit mass is independent of epoch, total mass and history.

In models with low $\lambda$, evolution and enrichment of the central parts are more rapid, hence the higher centre-to-edge difference in $\Upsilon_{\star}$ in all bands. A higher $v_c$ (and thus a higher galactic mass) results in higher $\Upsilon_{\star}$ at the centre due to older stellar populations caused by a more rapid infall rate of gas in higher density regions \citep{BP00,Heavens2004}. \\

\begin{figure}
\begin{center}
\epsscale{1}
\plotone{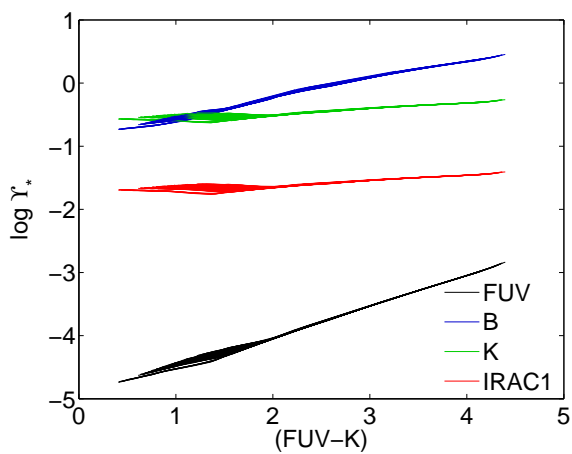}
\caption{\label{fig:UpsVscolour} $\Upsilon_{\star}$ vs. colour index for FUV, B, K, and IRAC1 as calculated from the ensemble of our models at every radius.}
\end{center}
\end{figure}

Our own relation between colour and $\Upsilon_{\star}$ for each band is presented in figure \ref{fig:UpsVscolour} and the coefficients of the first-degree polynomial in table \ref{tab:colourUpsParameters}, for the sake of comparison with  data from \citet{BdJ01}.

\begin{deluxetable}{ccccccccc}
\tabletypesize{\scriptsize}
\tablecaption{\label{tab:colourUpsParameters}Coefficients of the colour-$\Upsilon$ relation in different bands ($\log(\Upsilon_{\star}) = a_{\lambda}+b_{\lambda}(\mathrm{colour\,index})$)}
\tablewidth{0pt}
\tablehead{ \colhead{Colour index}&\colhead{a$_{FUV}$}	&\colhead{b$_{FUV}$}&\colhead{a$_{B}$}	&\colhead{b$_{B}$}&\colhead{a$_{K}$}	&\colhead{b$_{K}$} &\colhead{a$_{IRAC1}$}	&\colhead{b$_{IRAC1}$}
}
\startdata
B-R&-5.09&1.68 &-0.86 &1.01 &-0.69&0.30&-1.81&0.28\\
FUV-K&-5.24 &0.69 &-0.77 &0.35 &-0.68&0.10&-1.79&0.09\\
B-R\tablenotemark{a} &--- &--- &-1.33 &1.39 &-0.76 &0.47 &--- &--- \\
\enddata
\tablenotetext{a}{ From \citet{BdJ01}, scaled Salpeter IMF}

\end{deluxetable}

Our relation between $\Upsilon_{\star}$ and (B-R), while showing the same trend, has a significant offset from the most similar case which is the infall model of \citet{BdJ01} and also shows a slightly lower slope in both B and K bands.  This is due to the details of the ingredients of the two different models (the exact shape of the IMF, the star formation rate vs. gas density, etc.). Nevertheless, for the colour range covering galaxies (B-R $\sim$ 0.6-1.8), our relation in the optical agrees with that of \citet{Bell03}, showing that independent CSPE models yield similar results. 

A degeneracy of the relation causes a dispersion of the $\Upsilon_{\star}$ in  all  bands at very low colour index. This low index is found in the exterior regions of galaxies characterized  by populations with a more recent star formation history, a lower dispersion of ages and a higher fraction of young stars.  In these populations, the colour varies because of the contribution of young massive stars, but very old stars no longer dominate the luminosity and thus a variety of luminosities can correspond to the same colour index depending on the age dispersion of the population. These populations are characterized by a lower metallicity, which contributes to the scatter as well since colour-$\Upsilon_{\star}$ are less tight at low metallicities. It unfortunately means that in this particular colour regime, it is impossible to determine the $\Upsilon_{\star}$ unequivocally only by the observable colour gradient.

	\subsection{Mass models of individual galaxies}
	
\label{subsect:MassMod}

We present here the results of disc weighting by $\Upsilon_{\star}$ of all the individual galaxies of our sample as well as the mass modelling resulting for each.  Two different types of mass models were performed for each galaxy:
\begin{description}
 \item \textbf{a)} a conventional mass model where the disc's mass is inferred directly from the IRAC1 band photometry. Its $\Upsilon_{\star}$ and the dark halo parameters are let free to vary in a best-fit approach. 
  \item \textbf{b)} the median of the $\Upsilon_{\star}$-weighted discs in all available photometric bands is used as the stellar disc and the dark halo parameters are determined by a best-fit approach. 
 \end{description}    
 In the remainder of this paper, type-\textbf{a)} models are referred to as "constant-$\Upsilon_{\star}$ as free parameter" models and type-\textbf{b)} models are referred to as "weighted-disc" models.
 
A pseudo-isothermal halo was used for modelling in both cases, as was used for the scaling of the CSPE models.  The two mass models were performed by the interactive task manager \textit{rotmas} of the \textsc{Gipsy} package. 

The prominence of the bulge in the case of earlier type galaxies (NGC4321 and NGC4569) called for the break down of the photometry profile into bulge and disc components.  A conventional de Vaucouleurs bulge and strictly exponential disc were used \citep{deVauc48,Freeman70} with crude starting values set by fitting a straight line through the radius showing a visible agreement to an exponential disc and a subsequent  $\chi^2$ minimization approach.    

A $\chi^2$ fitting procedure was repeated on the exponential disc thus obtained to find the most appropriate model in the grid and the $\Upsilon_{\star}$ was calculated anew for those results. Contrary to the fit procedure adopted in JCM11, our procedure takes into account the fact that in those two particular galaxies, the UV photometry was either unavailable or unreliable so an inferior weight was given to UV and \textit{u} and \textit{g} bands compared to visible and IR bands.   The adopted $\Upsilon_{\star}$ conversion for the bulge  was supplied by the global relations connecting the colour index to the $\Upsilon_{\star}$ in each band presented in the previous section (relations shown in figure \ref{fig:UpsVscolour}).  \\

The case of barred galaxies cannot be ignored: it represents at least half of galaxies by conservative estimates \citep{deVauc63} and could reach as much as two-thirds of the population if the IR classification is considered \citep{Knapen2000,Hernandez2005}.  As was shown previously by numerous authors \citep{BournaudCombes2002,Athanassoula1992,Athanassoula2002, Hernandez2005}, the effect of bars on the dynamical potential is substantial, driving several processes to the point where a big part of the evolution of galaxies could be their doing \citep{KK04}. Not only does the bar modify the overall history of the galaxy, but it also alters the rotational velocities by transporting gas and stars towards the centre of the galaxy.  The alteration of the observed velocities thus depends upon the orientation of the bar with respect to the position angle of the galaxy \citep{Dicaire2008b}. If the bar is parallel to the major axis, the observed velocities are lower than would be expected from an axisymmetric potential; on the contrary if the bar is perpendicular to the major axis the observed velocities are higher. For the median case where both position angles display a 45 degrees angle, it does not have any effect whatsoever on the observed velocities \citep{Athanassoula2002}. The exact correction of a 2D velocity field would be beyond the scope of this paper, but it should  nevertheless be  possible to apply a coarse first-order correction to the rotation curve to palliate this known effect.  

\subsubsection{NGC 2403: the Typical Case}

Situated at a distance of 3.2 Mpc, NGC 2403 is the most typical late-type spiral in our sample. This galaxy does not show any sign of interaction even though it is part of the M82 group.  
\citet{Hernandez2005} studied this object in their BH$\alpha$Bar sample and concluded that although this galaxy is classified as an SAB, the bar is weak, not very well defined and has no notable effect on the kinematics. We thus present it first and use this galaxy as our ``textbook'' case to illustrate all steps of our method.    \cite{Dicaire2008} in their article on the SINGS H$\alpha$ sample measured the rotation curve up to a radius of 3.28~kpc  while the THINGS \textsc{Hi} data extends to 17.9~kpc. \citet{Blaisous2004} found some asymmetry in the rotation curve due to the presence of an arm starting at 200" (3.1~kpc). 

\begin{figure}
\begin{center}
\epsscale{1}
\plotone{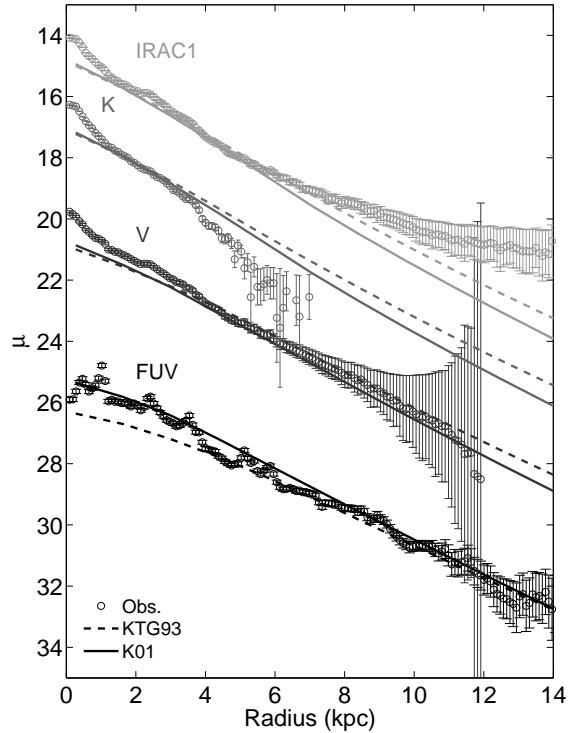}
\caption{\label{fig:NGC2403CompIMFPhoto}Fit of the disc models to the photometric data for IMFs KTG93 (dashed lines) and K01 (continuous lines).  Photometric curves have been offset for a clearer understanding: FUV +3mag, V (no offset), K -3mag, IRAC1 - 6mag. The 2.2$R_d$ of this galaxy is situated at 4.8~kpc.}
\end{center}
\end{figure}

Figure \ref{fig:NGC2403CompIMFPhoto} shows the fits to the photometry of the models with the two different IMF are equally good for both. The higher UV fluxes yielded by K01 models are clearly visible in the figure and in this particular case fit the observations better. As was already mentioned by JCM11, several galaxies show truncation or anti-truncation of the disc, which the models fail to reproduce as is clearly visible in the figure. However, since these discrepancies occur at a low signal-to-noise ratio and are within the uncertainties in the V-band, it is difficult to distinguish them from simple problems in the background subtraction. This departure from the classic exponential disc profile at outer radii has a very minimal effect on the overall mass model of the galaxy since these outer parts have a lesser impact on the global kinematics, and particularly on the determination of the halo parameters because very little mass is involved.

\begin{figure}
\begin{center}
\epsscale{1}
\plotone{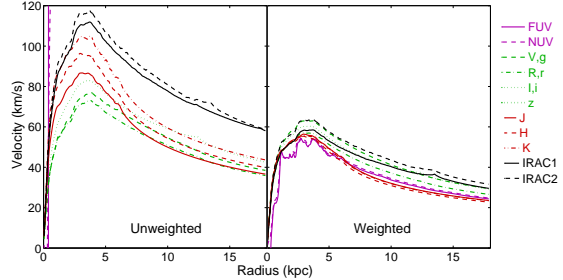}
\caption{\label{fig:NGC2403CompIMFVelocity} Rotation velocity of the discs of NGC2403 computed in different wavelengths from FUV- to IRAC-bands as a function of the galactic radius.  \textit{Left panel:} calculated by assigning an $\Upsilon_{\star}$ of 1 to the stellar density retrieved from the photometry profiles.  Shown for comparison purposes with the weighted disc models.  \textit{Right panel:} calculated from the photometry-derived density weighted by the modelled $\Upsilon_{\star}$. The scale does not allow FUV and NUV bands to be fully represented in the left panel as they reach $\simeq 10^3$km s$^{-1}$. }
\end{center}
\end{figure}

The weighting operation was performed with those two different sets of models (KTG93 and K01) on all galaxies, but since the fit gave slightly more consistent results for all galaxies with the latter, only the K01-weighted discs are presented here. In this particular case, both K01 and KTG93 models give good fits, even in the UV wavelengths, with differences in stellar surface densities of the order of $\approx 300 M_{\odot} \mathrm{pc}^{-2}$ at the centre when considering all bands. Figure \ref{fig:NGC2403CompIMFVelocity} shows the difference between the effective circular velocities of discs whose stellar density is derived directly from the photometry profiles (with $\Upsilon_{\star} = 1$) and discs whose stellar density is derived from the photometry profile and weighted by the CSPE models.

Due to the higher $\Upsilon_{\star}$ in the central regions of the galaxy, the velocity of the discs peak at lower radius than their unweighted counterpart.  This effect is all the more pronounced for galaxies represented by models with low $\lambda$ because of the much faster evolution of the inner parts of the galaxies.    
As a direct consequence, the mass models constructed with the weighted discs differ significantly from the ones derived from constant-$\Upsilon_{\star}$ as free parameter models.  As discussed in the previous section, infrared bands constitute an exception to this rule: their $\Upsilon_{\star}$ show less radial variation than the other bands to the point where it can almost be considered constant. In the example presented here, the best-fit model yields a halo with parameters ($\rho_0 =31.1 \mathcal{M}_{\odot}.\mathrm{pc}^{-3}$;$R_c = 3.6~\mathrm{kpc}$ and $\Upsilon=0.67$) for IRAC1 
unweighted disc compared to the ($\rho_0 = 142 \mathcal{M}_{\odot}.\mathrm{pc}^{-3}$;$R_c = 1.6 \mathrm{kpc}$) generated by the K01-weighted disc model. 
If we compare our results from the median disc with those available in the literature, we find we have very similar parameters of the dark halo to those of \citet{deBlokTHINGS} (theirs are $\rho_0 =153 \mathcal{M}_{\odot}.\mathrm{pc}^{-3}$;$R_c = 1.5~\mathrm{kpc}$). Their $\Upsilon_{\star}$ in IRAC1 band, originating from \citet{BdJ01}, is very different from the one we use for this band for the calculation of the median disc.  It is slightly more than twice ours but with more substantial radial variation.  

\begin{figure}
\begin{center}
\epsscale{1}
\plotone{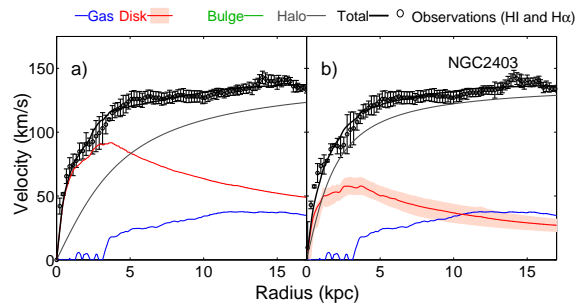}
\caption{\label{fig:NGC2403ModMass} Best-fit halo mass models of NGC2403 using: \textit{a)} $\Upsilon_{\star\,IRAC1}$-as free parameter mass model \textit{b)} CSPE-weighted median disc mass model.}
\end{center}
\end{figure}

\begin{figure}
\begin{center}
\epsscale{1}
\plotone{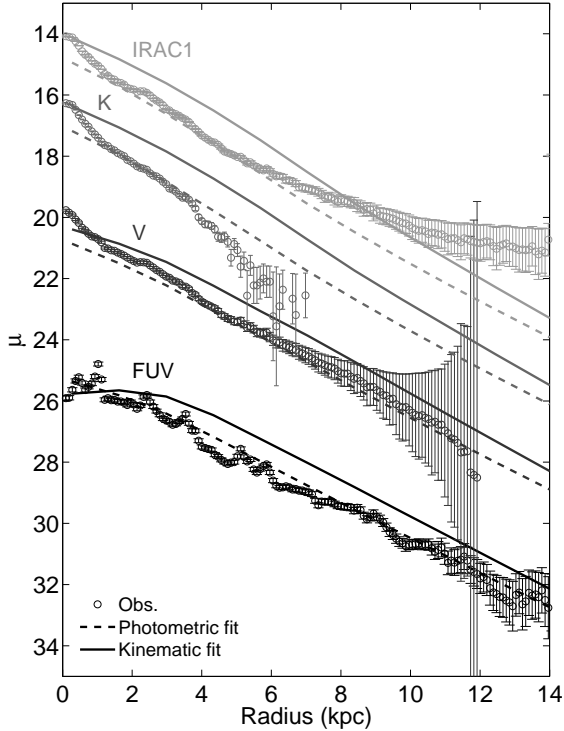}
\caption{\label{fig:ComparePhotosNGC2403}Comparison of the fits of CSPE models to the NGC2403 photometry: in dashed line, the former best photometric fit found by JCM11 ($\lambda=0.051$ and $v_c=116$) and in solid line the model corresponding to the best-fit halo of case b) in figure \ref{fig:NGC2403ModMass}}
\end{center}
\end{figure}

\begin{figure}
\begin{center}

\epsscale{1}
\plotone{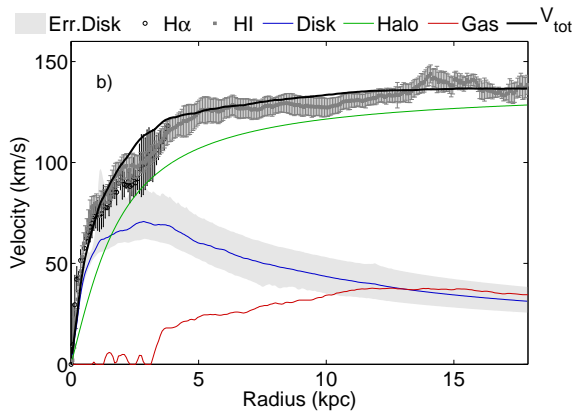}
\caption{\label{fig:MassModNGC2403KineFit} New mass model of NGC2403 with the weighting of the disc performed with the best-fit halo parameters of case b) in figure \ref{fig:NGC2403ModMass} and halo generated directly from those same parameters}
\end{center}
\end{figure}

A second iteration of this procedure was performed to take into account the kinematics as a supplemental constraint on the fit.  The parameters $R_c, \rho_0$ of the dark halo determined in weighted-discs mass models were converted in $\lambda$ and $v_c$ with the help of equations \ref{eq:HaloToModel} and the corresponding CSPE model was then used to start anew the weighting procedure.  In the case of this galaxy, the photometry of the new model is offset in magnitude, but it does reproduce the general trend in all bands (see figure \ref{fig:ComparePhotosNGC2403}).  The fit to the rotation curve in this new model is similarly good to the parameters of the weighted-disc model (see figure \ref{fig:MassModNGC2403KineFit}). 

This coherence does not warrant the veracity of this new solution and it should be taken with care: the photometry of the different bands weighted by the $\Upsilon_{\star}$ of this new simulation produce discs with greater variation from one photometric band to another than in the case of the one selected only by the match to the photometry of the disc.  The quadratic addition of all the velocity components in the mass model makes the more massive (i.e. the halo) component determine the overall appearance of the velocity curve and it therefore conceals the effect of the different weightings of the disc.   As we stated in section \ref{sect:Meth}, the exact form of the velocity curve has a lesser influence than its maximal velocity and thus the kinematics is a less constraining condition than the photometry on this type of models.    
  
In all other cases except for NGC 7793, the evolution model determined by the dark halo parameters found for our $\Upsilon_{\star}$-weighted disc model did not satisfactorily fit the photometric data and were not used to generate iterated mass models.

\subsubsection{Application of the Method to Other Galaxies}
\label{subsubsect:AllGalaxies}

Galaxies being as different as can be from one another, each one is described in detail in appendix to assess their particularities   while this paragraph provides a summary of the general tendencies of the whole sample.  \\

The unicity of each galaxy renders summaries difficult, but it is also a conclusion in itself.  The galaxies of our sample presenting either a bar or an enhanced stellar activity all showed less consistency between the mass density profiles derived from each band with one another compared to the ideal low-activity axisymmetric ones. All galaxies showed a convincing fit of the kinematics data when using CSPE models for the determination of the discs mass. In general, the best-fit parameters we evaluated from the $\Upsilon_{\star}$-weighted disc models were in good agreement with the results found in the most recent literature, notably with those published by the THINGS group except for a few galaxies for reasons detailed in the appendix.\\

Table \ref{tab:DHParameters} summarizes the results of the two different mass models performed for our whole sample of galaxies. The two last columns present the equivalent R$_c$ and $\rho_0$ of the scaling halo used in the evolution models. The conversion relations, from \citet{FallEfsthathiou80},\citet{BP99} and \citet{MMW98}
are the following:

\begin{eqnarray}\label{eq:HaloToModel}
R_c = \frac{70\lambda v_c}{220}\\
\rho_0 = \frac{v_c^2}{17.284\,\pi\,R_c^2}
\end{eqnarray}

\clearpage
\begin{deluxetable}{lc|ccc|ccc|cc}
\tabletypesize{\scriptsize}
\tablecaption{\label{tab:DHParameters} Disc's $\Upsilon_{\star}$ and dark halo parameters (pseudo isothermal sphere) for each galaxy in all three cases of mass models}
\tablewidth{0pt}
\tablehead{
\colhead{Galaxy}&\colhead{2.2R$_d$}		&\multicolumn{3}{c}{\small{$\Upsilon_{\star}$-as free parameter}}	&\multicolumn{3}{c}{\small{CSPE-determined $\Upsilon_{\star}$}}	&\multicolumn{2}{c}{\small{Scaling halo}}  \\
 &IRAC1	&\colhead{$R_c$}	&\colhead{$\rho_0$}	&\colhead{$\Upsilon_{\star\,IRAC}$}	&\colhead{$R_c$}	&\colhead{$\rho_0$}	&\colhead{$\Upsilon_{\star\,IRAC}$}	&\colhead{$R_c$}	&\colhead{$\rho_0$}	\\
	
	&\colhead{(kpc)} 	&\colhead{(kpc)}		&\colhead{($\times 10^{-3}\mathcal{M}_{\odot}\mathrm{pc}^{-3}$)}	&\colhead{} &\colhead{(kpc)}		&\colhead{($\times 10^{-3}\mathcal{M}_{\odot}\mathrm{pc}^{-3}$)}&\colhead{}	&\colhead{(kpc)}		&\colhead{($\times 10^{-3}\mathcal{M}_{\odot}\mathrm{pc}^{-3}$)}	
	}	 
   
\startdata
NGC925		&6.9		&5.20	&9.92	&0.01	&7.29	&11.5	&0.26-0.28	&4.59	&20 \\
NGC2403		&4.8		&3.55	&31.1	&0.67	&1.58	&142	&0.25-0.31	&1.88	&70\\
NGC3198		&6.1		&2.83	&47.1	&0.31	&2.38	&70.5	&0.27-0.37	&2.57	&82\\
NGC3621		&3.2		&5.20	&15.8	&0.34	&4.29	&26.6	&0.27-0.47	&1.16	&291\\
NGC4254		&5.2		&9.02	&18.1	&0.45	&10.60	&14.5	&0.28-0.49	&2.13	&232\\
NGC4321		&13.2	&1.27	&687.0	&0.29 (disc)&1.56	&395.8	&0.29-0.46	&3.73	&114\\
			&		&		&		&0.15 (bulge) &	&		&			&		&\\
NGC4569		&9.9		&8.08	&71.3 	&0.18 (disc)&25.7	&19.1	&0.40-0.46	&3.43	&114\\
			&		&		&		&0.09 (bulge)&		& 		&			&		&\\
NGC5055		&13.7	&3.63	&36.4	&0.60	&5.42	&18.9	&0.28-0.43	&2.38	&202\\
NGC7793		&2.3		&2.45	&40.3	&0.63	&1.23	&166.0	&0.25-0.32	&1.25	&120\\
DDO154          	&2.2		&1.57	&20.8	&		& 1.23	&30.6	&0.23-0.25	&0.78	&31\\
\enddata
\end{deluxetable}

\section{Discussion}
\label{sect:Discussion}

Let us discuss here further implications of the results presented so far and the limits of the method used.

A general remark should be given about the use of H$\alpha$ and H\textsc{I} data: although H$\alpha$ was formerly considered as providing the best resolution for kinematics data, H\textsc{I} data is now almost as spatially resolved.  Nevertheless, comparison between the two for a whole set of galaxies led to the conclusion that resolution is not the only factor at play to explain the differences.  \citet{DaiglePhD} concluded that the optical depth is to blame for the disparity commonly found between the two gaseous components. 

While we use bands from the NUV to the midIR, we are well aware that the UV is not the best band to estimate the stellar masses.  However, we decided to keep them because their impact on the median density is minimal.

 The set of CSPE models employed here is successful in reproducing a wide range of observable properties of galaxies. By assuming our Milky Way is typical, \citet{BP00} used the most precise data then available to calibrate several parameters (e.g. gas accretion rate and star formation efficiency) and further refined their method by taking into account some properties of a sample of nearby galaxies \citep{Boissier2003}. Ten years later with a plethora of data from large galactic surveys, \citet{JCM2011} verified that the model still predicts correctly the main characteristics of galaxies.  In the current study, we found that not only do those models reproduce well the photometry of galaxies, but they also supply physically motivated $\Upsilon_{\star}$ weighting factors leading to disc masses fluctuating very little from one photometry band to the other.  \\

 We believe our models provide good results to the first order in view of how well they reproduce the photometry profiles. In the future, some adjustments could make those models even more physically realistic. 
 First comes the question of the IMF.
 A universal IMF has been used here and seems to provide good results.  But while some authors consider the universal IMF to be representative \citep{ReviewIMF2010,Calzetti2011}, some others have raised the concern about its validity \citep{Meurer2009,Boselli2009}.  No unequivocal confirmation of this variability of the IMF has been supplied to date, but if it turned out to be the case, it would certainly be interesting to introduce those changes in the models. 
There also should be an update in the models for a better handling of the advanced stellar phases and circumstellar dust emission, especially in the NIR. Finally, radial transport should be implemented to consider cases (e.g. barred galaxies) where a significant mass exchange can take place. This could have complex effects on the UV profiles (lowering metallicity with outer gas, bringing fuel to the very centre) while the presence of bars might help in reducing star formation. Detailed investigations of such effects should then be performed.\\ 

We defined our "maximum-disc" as was originally meant in \citet{CF85}, i.e. as the maximum velocity that the disc can adopt without overshooting the observed velocity curve, and not as the 0.85$V_{max}$ fixed in \citet{Sackett97} because of the different conformation of the density profile of radially-varying $\Upsilon_{\star}$ and radially-constant $\Upsilon_{\star}$.

The approach used here, i.e. taking into account all photometric bands to construct a median stellar disc for mass modelling is very thorough but should not be necessary in view of the consistent results obtained for all bands in most galaxies.  The topic was discussed at length by several authors already \citep{BdJ01,BC03,deBlokTHINGS}, but let's stress once again that the infrared bands, and especially mid-infrared IRAC bands are indeed appropriate to determine disc masses because of their almost flat $\Upsilon_{\star}$ profile in the present day galaxies of the nearby universe.  In all galaxies, we found those bands to reproduce convincingly the median discs found by the full method.  The only difficulty is to find the appropriate CSPE model when one has only a few bands at its disposal.  The $\Upsilon_{\star}$-colour relation comes in handy in this case, but as one can see from figure \ref{fig:UpsVscolour} that no colour index allows for unambiguous determination of $\Upsilon_{\star}$ in the infrared bands at low colour index, though the best results would follow from the use of (FUV-R).  Only the outskirts of galaxies should be subject to such a problem and the mass model is less sensitive to variation in these regions than it would be if it occurred at more central radii where more mass is involved. \\  

It would be very interesting and most certainly worth additional work to establish the limits of the validity of these models for spheroidal components and then use them for spiral galaxies with important bulges or elliptical galaxies. Some effort should also be employed towards a method exploiting the full 2D information of the velocity maps. \citet{Zibetti2009} have already constructed 2D maps of $\Upsilon_{\star}$ of galaxies with the help of \citet{BdJ01} $\Upsilon_{\star}$-colour relationships.  This, along with analysis techniques of 2D velocity information, such as the ones developed for example by \citet{Wiegert2010}, is the next logical step leading to a realistic treatment of kinematics information in galaxies.\\

\begin{figure}
\begin{center}
\epsscale{1.1}
\plottwo{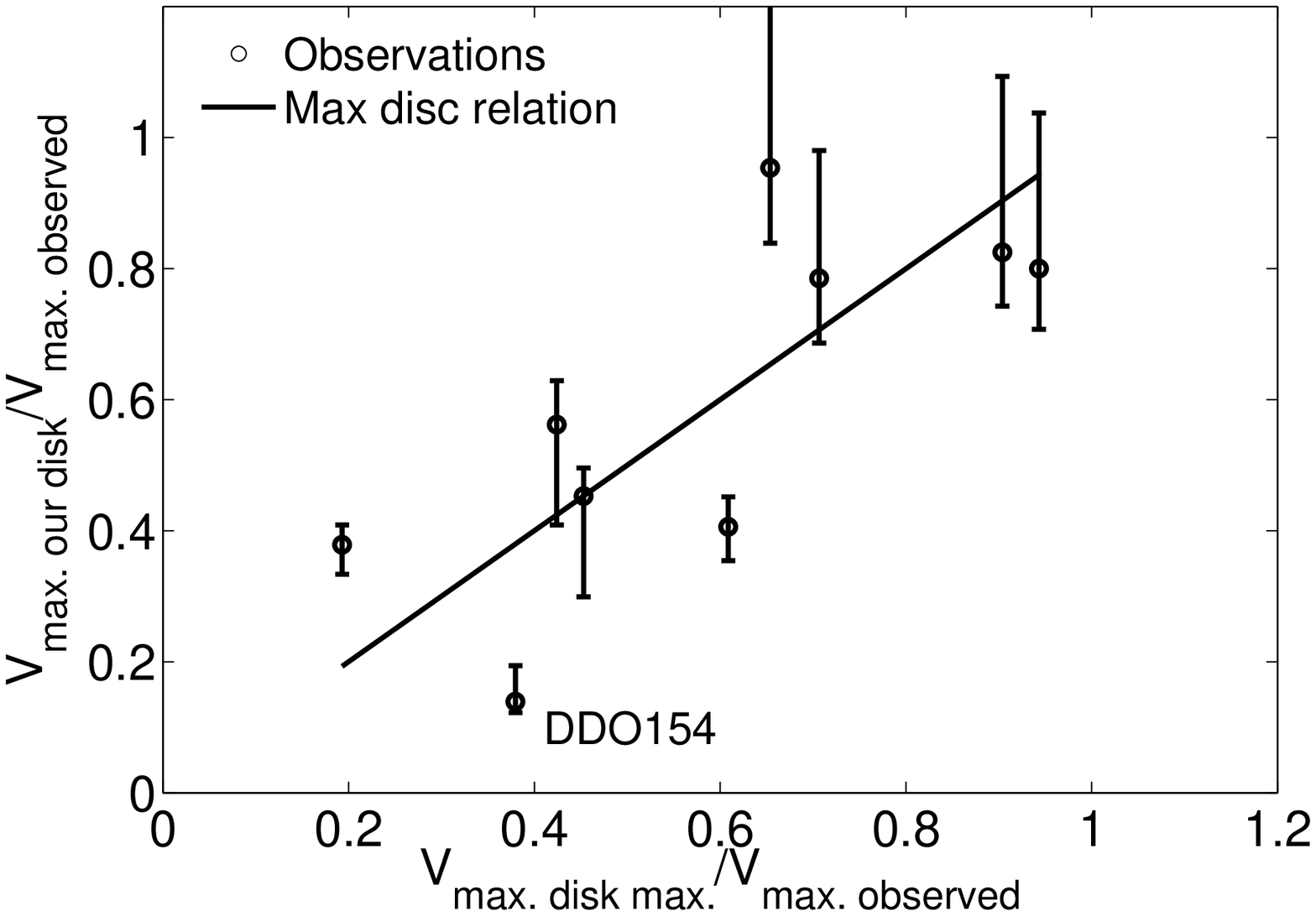}{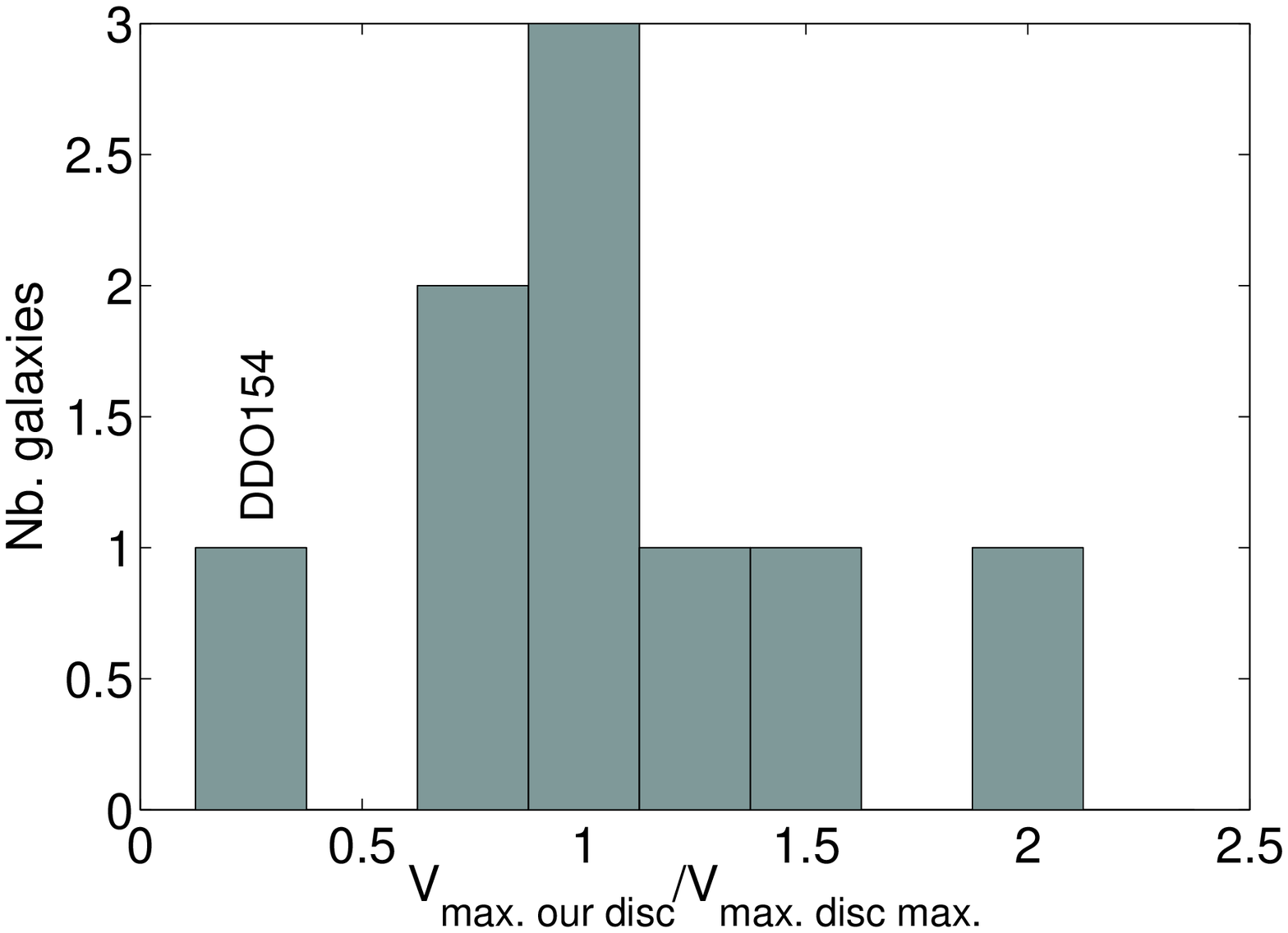}
\caption{\label{fig:TestMaxdisc}\textit{Left panel:} Comparison of the maximum velocity ratio between the maximum disc case and our case. \textit{Right panel:} histogram of the maximal velocity of our disc over the maximal velocity of a maximal disc.}
\end{center}
\end{figure}

Our discs are compatible with maximum discs because of their higher $\Upsilon_{\star}$ values in the centre.  This is ultimately due to the inside-out formation scheme allowing for more mass in the centre in the form of an older population. 
This conformity with the maximum disc hypothesis also means that in order not to overshoot the velocities in the centre all relevant corrections need be made on the rotation curve.\\
The individual results of the previous section show that elements such as a bulge or a bar can no longer be considered as insignificant with this new method of fixing the disc's mass.   It is important to mention the presence in figure \ref{fig:TestMaxdisc} of galaxies having discs that look over-maximal.   Even if at first glance it isn't physical to accept over-maximal discs, we kept them as is because the median disc was below the higher bound of the  error bars of the rotation curve, and that is not even considering the lower bound of the error bars on the mass of the disc. This nevertheless gives rise to questions about whether the stellar component should be systematically split up in bulge and disc components.  Figure \ref{fig:Decomposition} convincingly demonstrates the lowering effect of this breaking down on the disc's contribution to the rotation curve and galaxies with even a smallish bulge like NGC 3198 are probably tainted by the effects of a neglected bulge.      
\\

\begin{figure}
\begin{center}
\epsscale{1}
\plotone{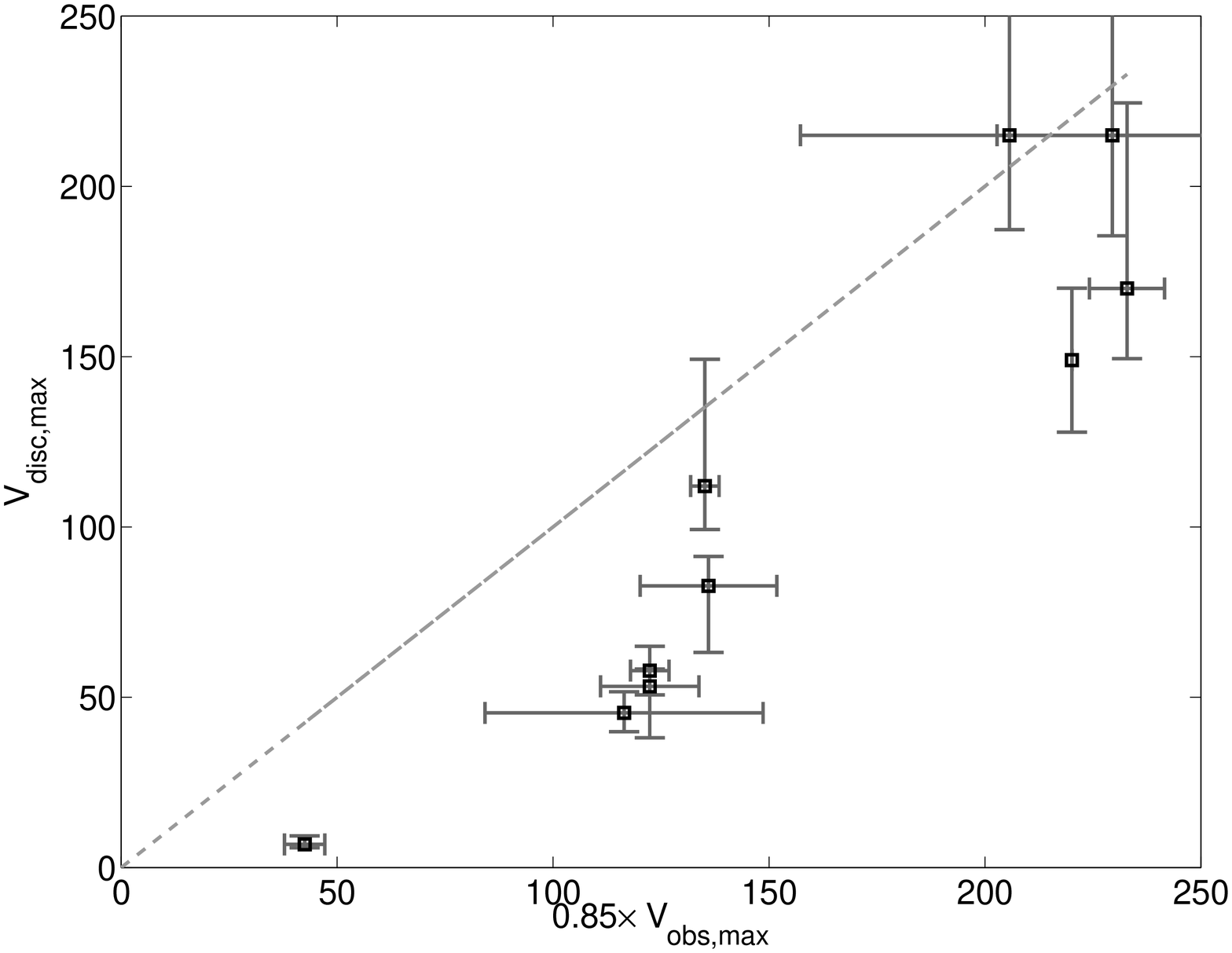}
\caption{\label{fig:TestMaxdiscSackett} Comparison of the maximum velocity of our discs with 0.85 of the maximal observed velocity of the galaxy. The light-grey dashed line represents the maximality criterion.}  
\end{center}
\end{figure}

We compared the maximal velocities of our discs with the maximal-disc hypothesis.  Figure \ref{fig:TestMaxdisc} should make it clear that maximum discs are compatible with our results.  Once again, this is due to the higher $\Upsilon_{\star}$ in the inner regions of discs, making the disc maximal at low radii. This is at odds with results from \citet{Bershady2011}, who observe sub maximal discs from a face-on sample of spiral galaxies. This is in part due to the definition they adopt of a maximal disc as being $V_{\mathrm{disc}_{\mathrm{max}}} = 0.85V_{\mathrm{max}}$ that does not take into account the amplified contribution due to the radially dependent $\Upsilon_{\star}$.  For the sake of reference with the literature, we traced the comparison between the maximal velocity of our discs with \citet{Sackett97} criterion for disc maximality (0.85$\times V_{obs,max}$). The results are shown in figure \ref{fig:TestMaxdiscSackett}. According to Sackett's criterion, all of our discs would be sub-maximal, but if we were to heighten our discs to reach $V_{disc,max} = 0.85V_{obs,max}$, the inner parts of the discs velocity would dramatically overshoot the actually observed rotation curves. Only one galaxy lies well below the max-disc relation: DDO~154, as was expected from previous studies of the mass distribution in dwarf galaxies. The smallest galaxies are dominated by the dark halo component and the stellar disc is still building up in our models. 

\begin{figure}
\begin{center}
\epsscale{1}
\plotone{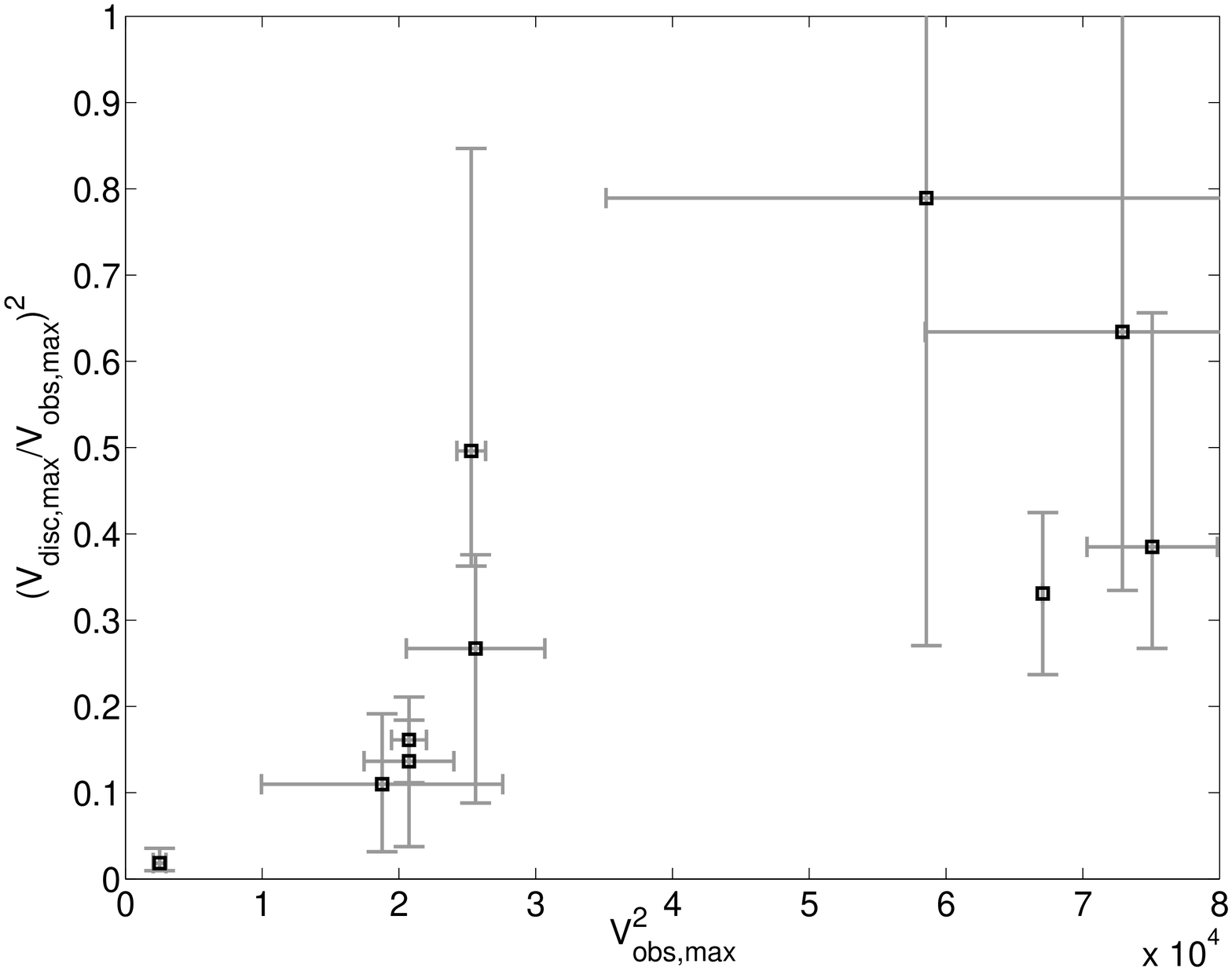}
\caption{\label{fig:TestProportion} Contribution of the disc to the overall mass of the galaxy as a function of the mass of the galaxy. } 
\end{center}
\end{figure}

Using velocities as the tracer of the mass, we also compared on a diagram the trend in mass of the disc proportion compared to the total mass of the galaxy ($M_{disc}/M_{galaxy}\propto{V_{max,disc}^2/V_{max,obs}}$). Figure \ref{fig:TestProportion} shows three distinct groups of galaxies: the higher mass group (NGC 4254, NGC 4321, NGC 4569 and NGC 5055), which shows a rather dispersed but nevertheless higher contribution of the disc to the overall mass, the middle mass group (NGC 925, NGC 2403, NGC3198, NGC3621 and NGC 7793), with a lower contribution of the disc, and the lower mass/dwarf galaxies group (DDO154), showing the lowest contribution of the disc to the overall mass. This is consistent with findings of \citet{Cote2000} and \citet{Donato2009} who find evidence of a trend of higher dominance by dark matter as the mass of galaxies decreases.

\section{Conclusions}
\label{sect:conclusion}

We presented in this article a method to infer the disc mass of galaxies from different photometric bands, from the NUV (Galex) to the MIR (IRAC-bands). This method has been applied to a sub-sample of ten SINGS galaxies. The conclusions we draw from this work are the following:

\begin{itemize}
\item The CSPE models prove to be a good tool to weigh the different photometric bands in order to obtain consistent stellar discs' masses regardless of the spectral band used.  The models provide radially dependent $\Upsilon_{\star}$.  The dispersion in effective circular velocity for all bands is on average of the order of $\sim$30 \%
\item Once the disc is determined by physically motivated models, it becomes impossible to ignore the effects of bulges and bars, so those essential corrections need to be made.
\item The agreement of the multi-wavelength observations with the model is a hint that a galaxy has had a standard evolution history and $\Upsilon_{\star}$ can be reliably applied even though each and every galaxy is unique and small discrepancies will inevitably arise.
\item Colour index vs. $\Upsilon_{\star}$ relation is an imperfect tool to assign masses to young stellar populations because of the degeneracy affecting $\Upsilon_{\star}$ in all bands at low colour index.
\item Mostly radius-independent mid-IR $\Upsilon_{\star}$ are advisable to use as the tracer of the stellar mass when only a limited number of photometry bands are available for present-day nearby galaxies.
\item Discs resulting from the method shown above are compatible with the max-disc hypothesis, and show a trend of higher disc contribution to the overall mass of the galaxy with increasing total mass of the latter.
\item For most galaxies, the halo used to perform the scaling of the properties of the model and the dark halo derived from the actual rotation curve agree to within 40\%.    
\item This method helped us achieve good results for both regular and dwarf galaxies.
\end{itemize} 
 
 A certain number of improvements can still be brought to it such as modifications to the CSPE models themselves and the full 2D treatment of kinematics data, but it nevertheless constitutes a step forward from constant-$\Upsilon_{\star}$ methods, mainly because the $\Upsilon_{\star}$ can no longer be considered as a free-parameter at any radius. 

Some interesting future work for this multi-wavelength method would be a similar study for higher redshift galaxies where the impact of younger stellar populations would significantly modify the appearance of $\Upsilon_{\star}$ profiles, particularly in the mid-IR.

\begin{acknowledgements}
We would like to thank Juan Carlos Mu\~noz-Mateos for kindly providing the results of his ready-to-use multi-wavelength data for all galaxies of our SINGS subsample.  We are also grateful to the THINGS \textsc{Hi} team for the availability of their data to the whole community.  

\end{acknowledgements}

\appendix
\section*{Appendix}

\paragraph{NGC 0925:}

The bar of this galaxy extends to 56.5"(2.55 kpc) and is oriented parallel to the major axis of the galaxy \citep{Martin1995,Hernandez2005}, hence its maximum effect on the rotation curve. As shown by \citet{Dicaire2008b} the true rotation velocity should be higher in the inner parts if proper corrections were applied for the presence of the bar. Accordingly, one can see on fig. \ref{fig:ModMassPanel1} a rotation curve behaving almost as a solid body and velocities of the K01-weighted discs being higher than the measured kinematics.  Our results are compatible with those from  \citet{Walter2008}.

\begin{figure}
\begin{center}
\epsscale{1}
\plotone{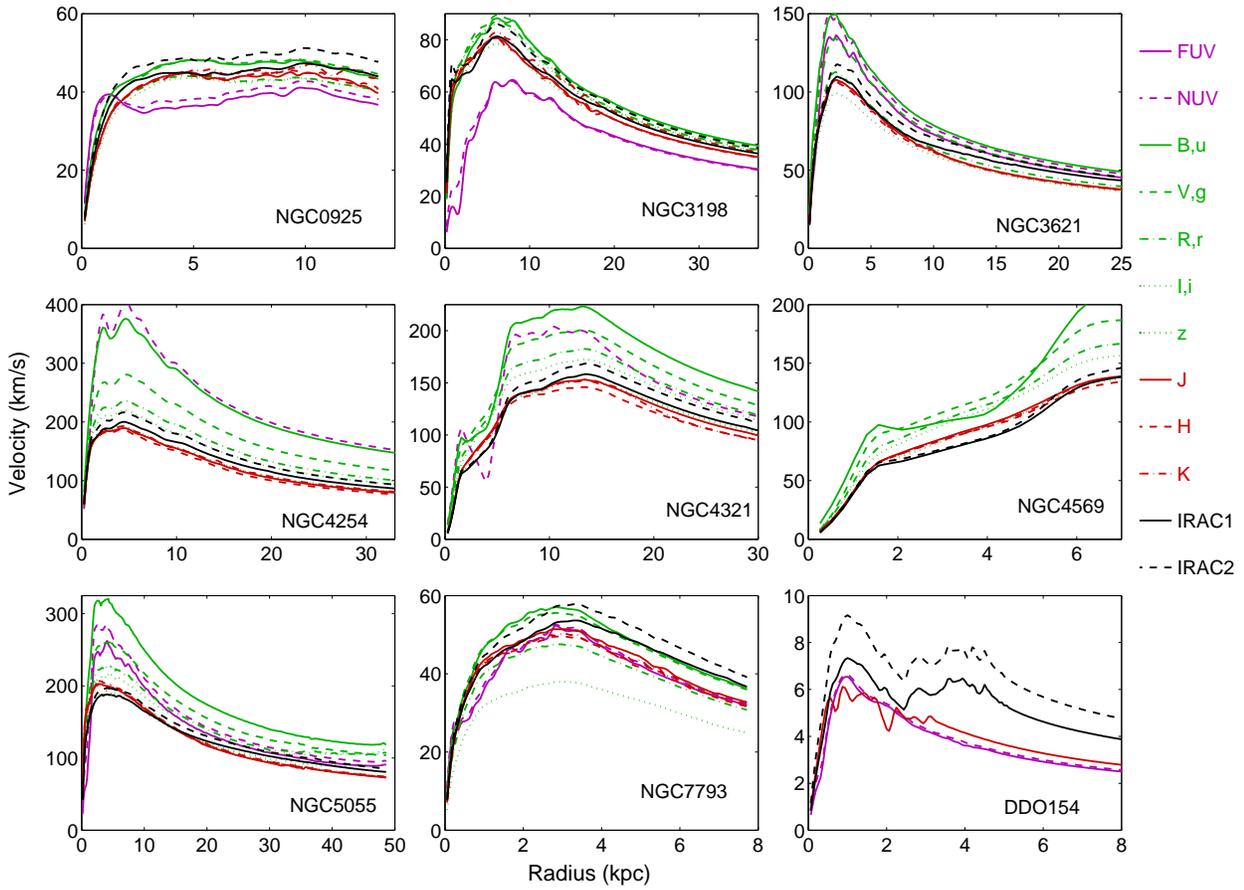}
\caption{\label{fig:AllDiscs} Rotation velocities of the discs for the whole sample of galaxies as seen in different wavelengths.}
\end{center}
\end{figure}

\begin{figure}
\begin{center}
\epsscale{1}
\plotone{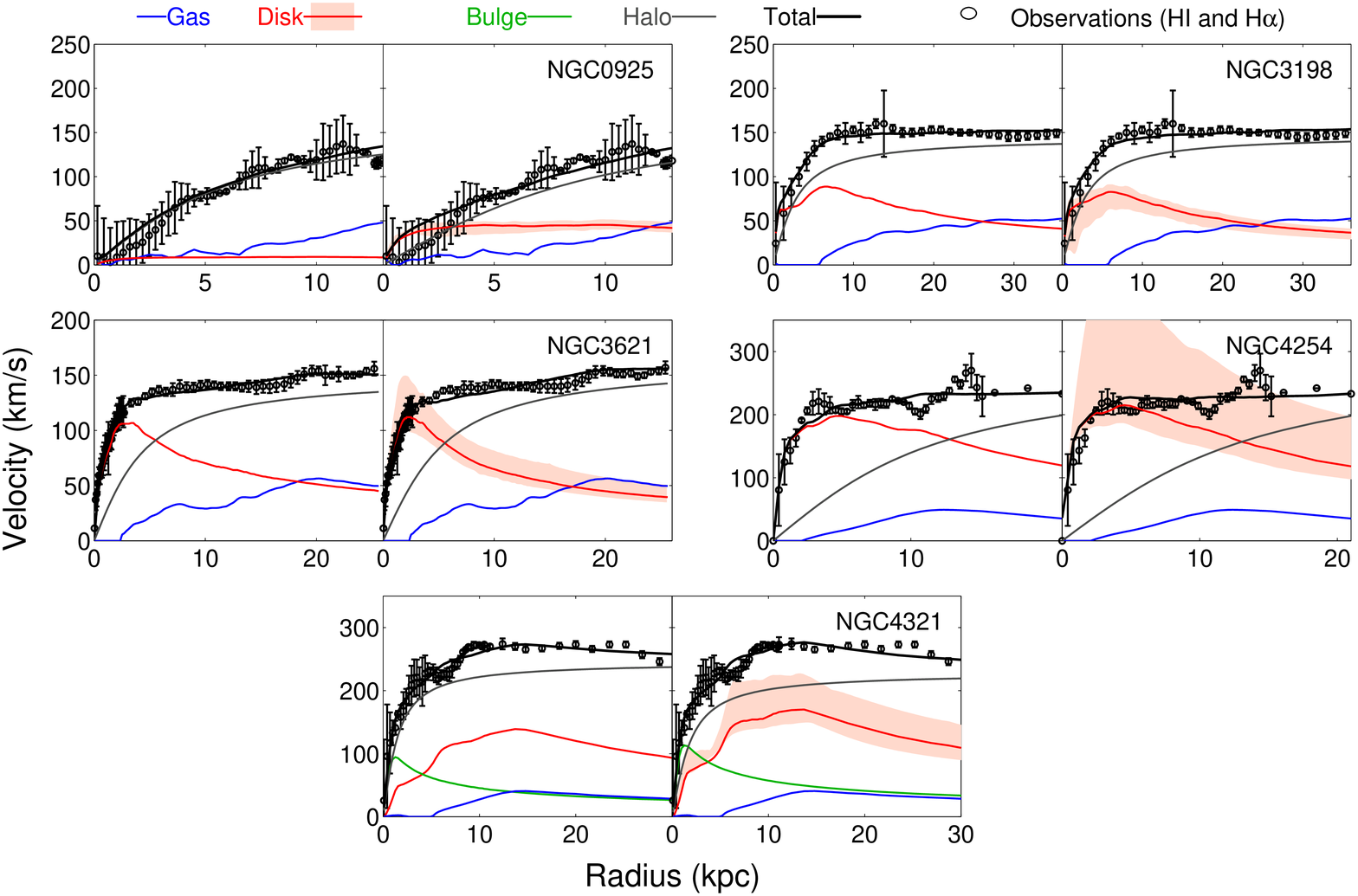}
\caption{\label{fig:ModMassPanel1}Mass models for the whole set of galaxies.  Left panels of each set is the model with $\Upsilon_{\mathrm{IRAC1}}$-as free parameter while right panels are models constructed from discs weighted by our models.}
\end{center}
\end{figure}

\begin{figure}
\begin{center}
\epsscale{1}
\plotone{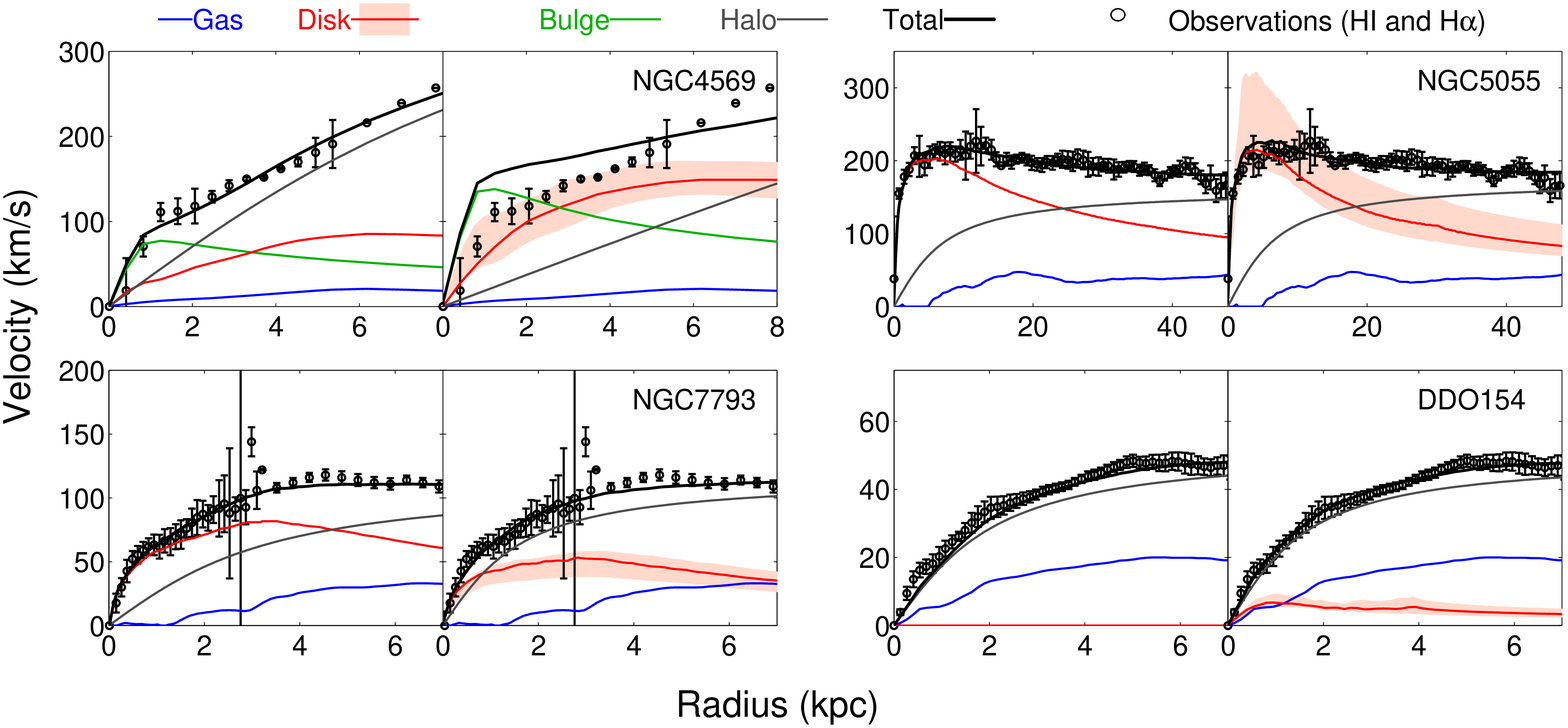}
\caption{\label{fig:ModMassPanel2}Mass models for the whole set of galaxies (continued).  Left panels of each set is the model with $\Upsilon_{\mathrm{IRAC1}}$-as free parameter while right panels are models constructed from discs weighted by our models.}
\end{center}
\end{figure}

\paragraph{NGC 3198:}
NGC 3198 is another representative of the barred family, albeit a lesser specimen.  The bar only extends to 81.8"(5.47~kpc) and its orientation is somewhat less detrimental to the rotation curve than in the case of NGC 0925 (P.A$_{gal}\sim 35^{\circ}$ and P.A.$_{bar}\sim 12^{\circ}$ ).  The rotation curve is nevertheless altered in the central regions and the agreement of the weighted stellar disc suffers from this situation as can be seen on figure \ref{fig:AllDiscs}. 
Because the velocity of the disc is contained in the error bars of the rotation curve, we still produced a model if only to compare with the abundant literature dedicated to this archetype of a spiral galaxy \citep{vanAlbada1985, Blaisous2001, deBlokTHINGS, Begeman1989}.  From the very start, NGC 3198 has been described constantly as symmetrical and regular.  Begeman, in order to produce a maximal disc, needed to adjust its $\Upsilon_{\star\,\mathrm{B}}$ as high as 3.8, which is far  from our $\Upsilon_{\star\,\mathrm{centre}}$ = 1.54. However, the strongest conclusion of \citet{vanAlbada1985} is that almost any combination of disc and halo masses yields convincing fits to the rotation curve (their results span $1.5\leq R_c \leq 12$~kpc and $4\times10^{-3}\leq \rho_0\leq 7.4\times10^{-3}$). \cite{Blaisous2001} support those conclusions and find equally good fit for different halo shapes.  Our values of $\Upsilon_{\star\,\mathrm{B}}$, $R_c$ and $\rho_0$ are best compared to those from \citet{deBlokTHINGS} because of the similarities in both our approaches and the fact that we used their HI data as they make it available for the community. Accordingly, as can be seen from table \ref{tab:ComparisonReference}, the best agreement is found with \citet{deBlokTHINGS} data.

\clearpage
\begin{deluxetable}{lllcc}
\tabletypesize{\scriptsize}
\tablecaption{\label{tab:ComparisonReference} Comparison of disc's and dark halo parameters from this work and other references}
\tablewidth{0pt}
\tablehead{
\colhead{Name}	&\colhead{Reference} 	&\colhead{$\Upsilon_{\star}$} &\colhead{$R_c$} &\colhead{$\rho_{0}$}	\\
	&	&	&\colhead{(kpc)}	&\colhead{$(\times 10^{-3} \mathcal{M}_{\odot}/\mathrm{pc}^3$)}	
}
\startdata
 NGC 0925	&This work			&0.26-0.28 (IRAC1)			&7.29	&11.5\\
 			&\citet{Walter2008}		&0.65 (IRAC1)				&9.67	&5.9\\
 &&&&\\
 NGC 2403	&This work			&0.25-0.31(IRAC1)			&1.58	&142.0\\
 			&\citet{deBlokTHINGS}	&0.30-0.60				&1.5		&153\\
			
 &&&&\\
 NGC 3198 	&This work 			& 0.21-1.54(B); 0.27-0.37(IRAC1) 	& 2.38 	& 70.5 \\
		     	&   \citet{vanAlbada1985} & 3.8 (B)					& 12 	& 4 \\
      			&  \cite{Blaisous2001} 	& 4.8 (B)					& 2.5 	& 5.7 \\
       			& \cite{deBlokTHINGS} 	& 0.7-0.8 (IRAC1) 			& 2.72 	& 47 \\
&&&&\\
NGC 3621 	&	This work 			& 0.27-0.47 (IRAC1)			& 4.29 	& 26.6  \\
   			&     \citet{deBlokTHINGS} & 0.4-0.8  (IRAC1)			& 2.8 	& 48.9  \\
  &&&&\\
NGC 4254	& This work			&0.3-0.53 (K)				&10.6	&14.5\\
			& \cite{Kranz2001} 		& 0.23-0.74 (K) 			& --- 		& --- \\ 
&&&&\\
NGC 4321	& This work			& 0.36-0.56(H)				&1.56	&395.8\\
			& \cite{Wada1998}		&0.2-0.8 (H)				&---		&---\\
&&&&\\			
NGC 4569   	&This work 			&0.40-0.46 (IRAC1)	  		& 25.71 	& 19.1  \\
       			
&&&&\\
NGC 5055	& This work 			& 0.31-0.42(undec.; IRAC1)	& 5.42 	& 18.9 \\
			&  \citet{deBlokTHINGS}	& 0.5-1.1(disc) \& 1.3(bulge) (IRAC1) & 45 	& 0.9\\
&&&&\\
NGC 7793	&This work 			& 0.23-0.84 (B)				&   1.23 	& 166\\
        			&\citet{CP90} 			& 2.2 (B)					& 2.7 	& 40 \\
        			&\citet{Dicaire2008}		& 2.6 (B)					& 2.9 	& 27 \\
&&&&\\
DDO 154		&This work 			& 0.16-0.22 (B), 0.23-0.25 (IRAC1)		&1.23 	& 30.6\\
        			&\citet{CP98} 			& 1.2 (B)					& 2.5 	& 22.0 \\
        			&\citet{deBlokTHINGS} 	& 0.32 (IRAC1)				& 1.32 	& 28.5 \\
\enddata
\end{deluxetable}
  
\paragraph{NGC 3621:}
For this late type galaxy, the CSPE model gives less satisfactory fits in B and V bands, yielding  discs whose mass diverge from the one determined by other bands.   The halo we find shows a core radius 3 times larger and a central density 4 times less concentrated than the one of \citet{deBlokTHINGS} because the surface density or our disc drops a lot more rapidly than theirs (R$>2.5$~kpc).

\paragraph{NGC 4254 (M99):}
This galaxy (as well as the following two) is part of the Virgo cluster and accordingly shows signs of past interaction. It is largely known for its visual asymmetrical aspect due to spiral modes m=1,2,4.   According to \citet{Chemin2006,Guhathakurta1988}, this very bright spiral possesses an asymmetric velocity field with streaming motions along the spiral arms. The asymmetry could be due to gas accretion according to \citet{Phookun1993} or to rapid and violent tidal interaction followed by ever-increasing ram pressure stripping according to \citet{Vollmer2005}.   \citet{Haynes2007} have investigated a scenario of galaxy harassment and found it plausible for NGC 4254 due to its high systemic velocity, long (242~kpc) H\textsc{I} tail and distance ($\sim$1~Mpc) from the centre of the cluster. Figure \ref{fig:AllDiscs} shows the very large dispersion of final discs masses from the UV to the IR when weighted by our $\Upsilon_{\star}$.  Obviously, NUV, u and g bands $\Upsilon_{\star}$ are overestimated.  This indicates that the models don't reproduce correctly the star formation history as a function of radius and it is probably due to interaction. The model in this case constrain poorly $\Upsilon_{\star}$ and its radial variation. 

The case \textbf{b)} mass model provides a passable fit to the data. It is not very surprising, due to the non-circular effects above mentioned, that the model would fit well the general trend but not the details of the rotation curve. Using different estimators, \cite{Kranz2001} find $\Upsilon_{\star\,K}$ consistent in the mean with ours ($\Upsilon_{\star\,K} = $ 0.23-0.74 compared to our $\Upsilon_{\star\,K}\sim$0.53).

\paragraph{NGC 4321 (M100):}

Another member of the Virgo cluster, this galaxy with noteworthy arms and bulge is one of those requiring bulge/disc decomposition and subsequent refitting to find the most appropriate model.  In order to split the two components, we performed a least-square fit of the photometry profiles by a sum of a pure de Vaucouleurs spheroid and an exponential disc \citep{deVauc63,Freeman70}.  The first guess estimate was provided by fitting an exponential disc to the region affected neither by the bulge nor the arms and subtracting the component from the photometry profile to determine the parameters of the spheroidal component.  The size of the bulge was let free to vary in all bands, but the result gave consistent size estimates.
Only the disc was used to find the best model in the grid. 
This new fitting procedure is delicate due to the peculiar photometry profile caused by the presence of the arms between 4.4-6.1~kpc and 8.7-12.2~kpc. We thus performed the new $\chi^2$-minimization excluding the affected radii.     
The quality of the fit might have suffered from taking into account only a subset of the available radii.  

We did not adopt a radius-dependant $\Upsilon_{\star}$ for the bulge since the stellar population in this case is, by its evolution history and its dynamics, thought to be much more uniform than the one in the disc.  This of course might be disputed when a pseudo-bulge is considered but should be correct for a typical de Vaucouleurs spheroid. Indeed, we verified that neither of our two earlier-type galaxies showed a strong colour gradient in their bulges before settling the issue.  The density we adopted for the bulge comes from data of the extracted bulge component in H-band weighted by the composite colour-$\Upsilon_{\star}$ relation presented in section \ref{sect:UPS}.  

The difference between the as-is photometry curve and the quadratic sum of its bulge and disc components is striking ($\Delta v \approx 75$~km s$^{-1}$), especially in the innermost radii where the fate of the mass model is sealed. This grand design spiral, just like many members of the Virgo cluster, has been studied by several authors.  \citet{Knapen1993} remark that the curve is still rising at their last radius and asymmetries are easily visible all the way from 3' to the exterior.  

\begin{figure}
\begin{center}
\epsscale{0.6}
\plotone{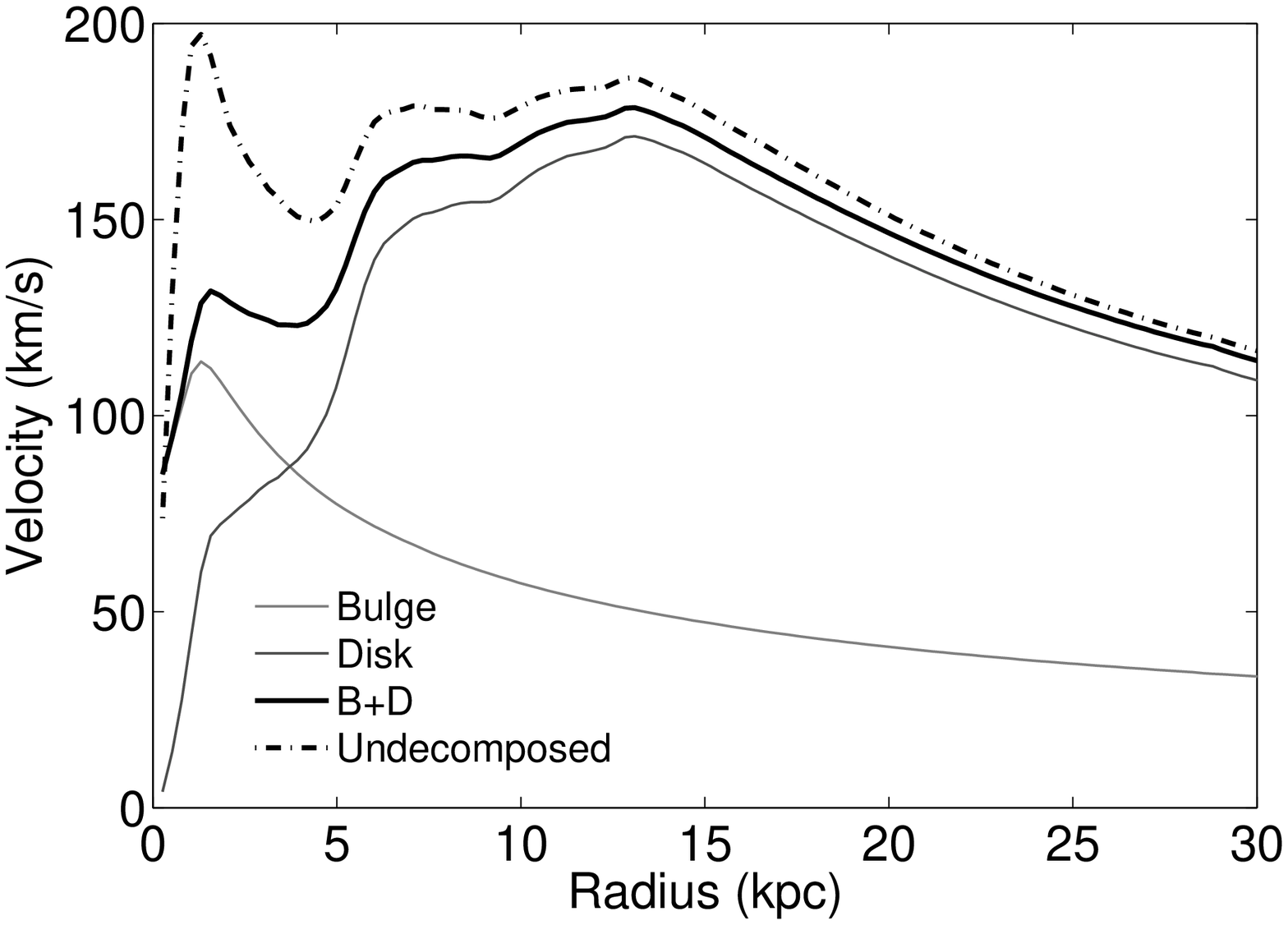}
\caption{\label{fig:Decomposition}Computed velocity of the stellar photometry of NGC4321. Respective velocities of as-is photometry and decomposition of the stellar photometry into bulge and disc components are compared.  The discs used are the median discs and the bulge is evaluated from the IRAC1 band.}
\end{center}
\end{figure}

NGC 4321 has a $\sim$60" ($\sim$5~kpc) bar in its centre whose pattern speed  has been characterized using the Tremaine-Weinberg relation by \citet{Hernandez2005} who find an $\Omega_p = 30.3$~km s$^{-1}$.kpc$^{-1}$. This bar is oriented perpendicular to the major axis, which should have the effect of raising the rotation curve (just the opposite of the situation in NGC 925).  \citet{Knapen1993} also discuss this bar and measures an increase in the circular velocity of as much as 50~km s$^{-1}$ at the extremity of the bar.   This bar might be responsible for the slight bump visible in the disc component in figure \ref{fig:Decomposition} even after a careful bulge+disc breakdown of the velocity profile. 
It is very hard to tell apart the rising of the curve due to the bar from the non-circular effect of the very strong arms, just as it is hard to distinguish the effect of the bulge and the bar on the luminosity profile. But if we applied a velocity correction of such big amplitude as the one suggested by \citet{Knapen1993}, it would clearly be difficult to reconcile the observed velocity with the one due to the combined bulge and disc.   \cite{Wada1998} find an $\Upsilon_{\star\,H}$ ratio which is consistent with our own estimate at the very centre.

\paragraph{NGC 4569 (M90):} 

This galaxy is far from being a textbook case for kinematics studies. Not only does it possess a considerable bulge, but its position in the Virgo cluster and its well-documented gas depletion makes it a weird beast of the galactic zoo \citep{Vollmer2004, Boone2007, Chemin2006}.  As such, a special model was devised taking into account the gas-stripping event from ram pressure leading to the peculiar SFH of the galaxy to compute the stellar disc density \citep{Boselli2006}.  As was the case for NGC 4321, the splitting of the profile in bulge and disc constituents results in a considerable change in the effective velocities.  This, of course, is due to the spheroidal rather than planar distribution of the mass in a bulge.   

The agreement of the rotational velocity of the weighted stellar component (disc and bulge) to the actual measured H$\alpha$ and \textsc{Hi} rotation curves is visibly inadequate in the inner regions where the bulge dominates the contribution of said stellar component.  Let us stress that the BP99 and BP00 models describe the evolution of disc galaxies and not of spheroidal object. The $\Upsilon_{\star}$ computed from these models can therefore be off because of differences in the star formation histories of bulge and disc.  The almost solid body appearance of the RC could be due to a past stripping event according to \citet{Vollmer2004}.  Another reason for this discrepancy might be the large-scale bar supposedly present in the centre, where a lot of emission due to a strong star-forming nucleus can be seen \citep{Boone2007, Laurikainen2002}. This bar sits at an angle compatible with a decrease in the observed rotational velocities.

\paragraph{NGC 5055 (M63):}

NGC 5055 is a moderately inclined SAbc galaxy with a slightly declining rotation curve.  \citet{Blaisous2004} found strong kinematic motions in the inner regions ($R<300$~pc).  The dark halo parameters we obtain are radically different from those of \citet{deBlokTHINGS} because they performed a bulge/disc decomposition and we did not.  One might think that in this case the dark halo we find should be less concentrated than the THINGS team because by not splitting the luminosity profile we overestimate  the rotational velocities of the disc.  The situation is however a little different, the $\Upsilon_{\star}$ used by \citet{deBlokTHINGS} being higher than ours.

\paragraph{NGC 7793:}

This galaxy could also have been used as a template model for our studies but its slightly higher inclination made us prefer NGC2403.  The models fit well the photometry in all bands and accordingly the error on the determination of the disc is low.

\citet{Dicaire2008} showed the rotation curve to be truly declining with their very deep H$\alpha$ observations reaching the confines of the THINGS rotation curve.   They used the B-band photometry to estimate the disc's mass and an isothermal dark halo.  
As can be seen in table \ref{tab:ComparisonReference}, the parameters of our isothermal halo and those from \cite{Dicaire2008} and \cite{CP90} are quite different because of the dissimilar values chosen for  our $\Upsilon_{\star\,B}$ and theirs. Because of our much lower value of $\Upsilon_{\star\,B}$ and its behaviour of rapidly decreasing with radius, we therefore find a more concentrated halo with higher central densities.

\paragraph{DDO 154:}

We chose DDO 154 as a test of our method on dwarf galaxies. 
It is not a benign question because of the overlook of radial transport and outflows in the models.  While this omission has negligible effects in "regular galaxies" it was suspected to have more perceptible repercussions on dwarf galaxies.  Models supplementing the original grid were calculated for $v_c = [20..50]$ in the $\lambda=[0.06..0.08]$ range.  The $\chi^2$ fitting procedure was once again performed on this grid extension and the result reproduced DDO 154's photometry as well as it had the regular galaxies. Quality photometry data was available only for five bands, but those included UV and IR bands.  We stick to performing the mass models with the K01 models-weighting even though in this particular case the KTG93 models reconciles better the UV and IR ranges.     
The mass model presented in figure \ref{fig:ModMassPanel2} shows the overall domination of dark matter at all radii of the galaxy. The disc's contribution to velocity lies far below the total rotation curve.  As can be seen on table \ref{tab:ComparisonReference}, our results differ from those of \citet{CP98}. Our $\Upsilon_{\star}$ is $\sim$6 times lower than theirs, thus our halo is two times more centrally concentrated than theirs while the central concentration is not very different.  On the other hand, if we compare our results to \citet{deBlokTHINGS} who follow a similar procedure to our own, we find almost perfect agreement.

\end{document}